%% file: paper.tex
\documentclass[letterpaper,twocolumn,10pt]{article}
\usepackage{usenix,epsfig,endnotes}
\usepackage[T1]{fontenc}
\usepackage{booktabs}
\usepackage{xurl}
\usepackage{xspace}
\usepackage[toc,titletoc,title]{appendix} 
\usepackage{colortbl} 
\usepackage{subcaption}
\usepackage{graphicx}
\usepackage[table,xcdraw,usenames,dvipsnames,svgnames]{xcolor}
\usepackage{xpatch}
\usepackage{multirow}
\usepackage{amssymb}
\usepackage[symbol]{footmisc}
\usepackage{gensymb}
\usepackage{pgfplots}
\usepackage{layouts}
\pgfplotsset{compat=newest}

\usepackage[backend=biber,style=trad-abbrv,maxcitenames=2,hyperref=true,
    url=false,
    isbn=false,
    backref=true,
    doi=false,
    eprint=false, bibencoding=utf8]{biblatex}
\usepackage{comment}
\usepackage[inline]{enumitem}
\usepackage{hyperref}
\usepackage[noabbrev, nameinlink]{cleveref}
\definecolor{raven}{RGB}{55,61,63}
\definecolor{resumeblue}{RGB}{13,61,86}
\definecolor{linkcolor}{rgb}{0.65,0,0}
\definecolor{citecolor}{rgb}{0,0.4,0}
\definecolor{urlcolor}{rgb}{0,0,0.65}
\definecolor{Gray}{gray}{0.9}
\hypersetup{
	pdfpagemode=UseNone,
	pdfstartview=FitH,
	colorlinks=true,
	linkcolor=linkcolor,
	urlcolor=urlcolor,
	citecolor=citecolor,
	bookmarksnumbered,
	breaklinks,
	bookmarksopen=false,
}

\DeclareNameAlias{default}{family-given}

\AtBeginBibliography{%
}

\DefineBibliographyStrings{english}{%
 andothers = {\addcomma\addspace\textsc{et\addabbrvspace al}\adddot},
 and = {\textsc{and}}
}

\DeclareFieldFormat
  [article,inbook,incollection,inproceedings,patent,thesis,unpublished]
  {title}{#1}

\renewbibmacro{in:}{%
  \ifentrytype{article}{%
  }{%
    \printtext{\bibstring{in}\intitlepunct}%
  }%
}

\renewbibmacro*{volume+number+eid}{%
  \printfield{volume}%
  \setunit*{\addcomma\space}%
  \printfield{number}%
  \setunit{\addcomma\space}%
  \printfield{eid}}

\DeclareFieldFormat{pages}{#1}

\renewbibmacro*{publisher+location+date}{%
  \printlist{publisher}%
  \setunit*{\addcomma\space}%
  \printlist{location}%
  \setunit*{\addcomma\space}%
  \usebibmacro{date}%
  \newunit}

\bibliography{paper.bib}

\definecolor{raven}{RGB}{55,61,63}
\definecolor{resumeblue}{RGB}{13,61,86}
\definecolor{linkcolor}{rgb}{0.65,0,0}
\definecolor{citecolor}{rgb}{0,0.4,0}
\definecolor{urlcolor}{rgb}{0,0,0.65}
\definecolor{Gray}{gray}{0.9}

\usepackage{soul}
\makeatletter
\patchcmd{\SOUL@ulunderline}{\dimen@}{\SOUL@dimen}{}{}
\patchcmd{\SOUL@ulunderline}{\dimen@}{\SOUL@dimen}{}{}
\patchcmd{\SOUL@ulunderline}{\dimen@}{\SOUL@dimen}{}{}
\newdimen\SOUL@dimen
\makeatother

\usepackage{pifont}

\newcommand{\parhead}[1]{\vspace{0.6pt plus 1pt minus 1pt}\par\noindent\textbf{#1}\hspace{.70em plus .5em minus .5em}}

\pdfoutput=1
\usepackage[
			activate={true,compatibility},
			final,
			tracking=true,
			factor=1100,
			stretch=10,
			shrink=10]{microtype}

\begin{document}

\date{Version May 2024}

\title{\Large \bf Watching the watchers: bias and vulnerability in remote proctoring software }

\author{
  {\rm Ben Burgess}\\
  Princeton University
  \and
  {\rm Avi Ginsberg}\\
  Georgetown Law
  \and
  {\rm Edward W. Felten}\\
  Princeton University
  \and
  {\rm Shaanan Cohney}\\
  University of Melbourne
}
\maketitle

\thispagestyle{empty}


\begin{abstract}
  Educators are rapidly switching to remote proctoring and examination software for their testing needs, both due to the COVID-19 pandemic and the expanding virtualization of the education sector.  State boards are increasingly utilizing these software for high stakes legal and medical licensing exams. Three key concerns arise with the use of these complex software:  exam integrity, exam procedural fairness, and exam-taker security and privacy.

  We conduct the first technical analysis of each of these concerns through a case study of four primary proctoring suites used in U.S. law school and state attorney licensing exams.  We reverse engineer these proctoring suites and find that despite promises of high-security, all their anti-cheating measures can be trivially bypassed and can pose significant user security risks.

  We evaluate current facial recognition classifiers alongside the classifier used by Examplify, the legal exam proctoring suite with the largest market share, to ascertain their accuracy and determine whether faces with certain skin tones are more readily flagged for cheating.  Finally, we offer recommendations to improve the integrity and fairness of the remotely proctored exam experience.
\end{abstract}

\normalsize
\section{Introduction}

The rapid adoption of proctoring suites (software for monitoring students while they take exams remotely) is a phenomenon precipitated by the pandemic \autocite{raman2021adoption}. Subsequently, we find high stakes exams being administered remotely.  Recent media coverage provides anecdotal evidence that these proctoring suites introduce security concerns, harm procedural fairness,  and pose risk to exam integrity \autocite{bloombergbias,nytimesdartmouth}. In this work, we aim to investigate the anecdotal claims and systematize the study of these software. We evaluate the proctoring software on its technical merits using traditional reverse engineering techniques and perform a fairness analysis.

Historically, computerized test taking benefited from in-person proctors and institution controlled hardware.  In the remote test-taking setting such proctors are unavailable so institutions resort to utilizing proctoring suites.  Proctoring suites attempt to mitigate the increased risks of cheating associated with the remote environment and student hardware by installing pervasive, highly privileged services on students’ computers.  Such software attempts to reduce academic misconduct during exams by limiting access to online resources and system functions, such as the ability to paste in pre-written materials. Considerable privacy concerns arise since a student's laptop is not generally a single use device that only contains class related material. A student using their own hardware faces the risk of their personal information being misused or leaked by the proctoring software. 

Initially, the remote proctoring software was marketed and designed for tertiary institutions, however, the software has recently been adopted for medicine and law licensing exams.  Inherent societal costs to illegitimately passing students are magnified in both professions, where an inept lawyer can put an individual's liberty at stake and an incompetent physician can cause significant trauma to patients.  The time and monetary burden of professional education and licensing places extreme pressure on students and this, along with the benefits of passing, increases the extent to which students may be willing to risk cheating. Maintaining confidence that degrees earned and licenses obtained ensure a minimum degree of competency and knowledge is imperative.  Equally important is the confidence that no individual who merits entrance into a profession has been blocked due to false cheating allegations.  Applied research into whether remote exam proctoring puts either of these interests in jeopardy is lacking in current literature and merits attention from the security and privacy community.

We conduct the first systematic, technical analysis of the remote proctoring ecosystem of law school and state bar exam boards.  By limiting the scope of our investigation to the single regulated profession of law, we can map out how the software is used across the entire profession, increasing confidence in our proposed recommendations.  We do not, however, have reason to suspect that applying our methodology to other examination settings would yield substantially different results.

Through public data aggregation and survey, four exam proctoring suites--—Examplify, ILG Exam360, Exam4, and Electronic Blue Book—--utilized in 93\% of U.S. law schools and 100\% of remote state bar exams are identified.  Reverse engineering of these four proctoring suites reveals vulnerabilities of varying complexity that compromise the purportedly secure testing environments. We evaluate suites in the context of three potential adversaries: a law student; a law student with computer science experience; and an experienced reverse engineer and discuss vulnerabilities identified with each. We find the majority of proctoring suites analyzed install a highly privileged system service that has full access to a user's activities.  Highlighting the privacy trade-off of this service, we find that system logs created \textit{before an exam begins} are nonetheless transmitted by the software to the vendor's servers during the exam.

We determine that Examplify implements a facial recognition classifier to authenticate a student against a pre-existing photograph prior to starting an exam. The classifier is then re-run repeatedly during the exam depending on the settings selected by the exam administrator. At each interval, the classifier attempts to determine whether the student initially authenticated is the student who continues taking the exam.  We extract the name of the facial recognition system Examplify is using, `face-api.js', and note they employ the pretrained models that are publicly available on the face-api's GitHub.   We test these models against current off-the-shelf state-of-the-art (SOTA) classifiers across several different datasets to evaluate whether one is more accurate than the other but find significant accuracy concerns across the board.

We then evaluate the classifiers across subjects of different races and find concerning variability. We investigate proctoring suite terms of service and user interface to shed light on whether a user is giving informed consent to all the monitoring.

We conclude with privacy protecting recommendations for remote exam proctoring administrators and students and provide vendors with suggestions to improve exam integrity while lessening the student privacy impact.

\smallskip

\parhead{Contributions}
\begin{itemize}
  \item We survey the top 180 law schools and all U.S. state bar associations to determine their remote proctoring practices (dataset to be released upon publication) (\Cref{sec:survey}).
  \item We reverse engineer four exam proctoring suites and identify the security the proctoring suite provides the institution and its student privacy ramifications (\Cref{sec:security}).
  \item We release a tool for automatically and rapidly extracting key security and privacy properties of exam suite software.
  \item We perform a detailed evaluation of racial biases in the facial recognition model used in the software with the dominant market-share for remote proctoring in law schools (\Cref{sec:fairnessaccuracy}).
\end{itemize}

While we acknowledge that the attack techniques we use are in general, well known, their application to the particular high-stakes context merits a comprehensive analysis.

\parhead{Research Ethics \& Limitations}
\smallskip

While our survey involved contacting law schools and state bar associations to determine what platforms they use and how they use them, our work was exempt from IRB review as we did not collect data pertaining to individuals.

Our analysis of facial recognition systems used a data set containing images of incarcerated individuals who were not given an opportunity by the original researcher to consent to its use for research. We consulted with an applied ethicist independent from the project and determined that while we are unable to rectify the consent issue to fully uphold the principle of autonomy, use of these images is nonetheless appropriate as our work fulfills the principle of beneficence in the following manner:

\begin{itemize}
  \item Our work aims to aid marginalized groups and dis-empowered individuals by evaluating the privacy/bias concerns of software that powerful organizations require students to use.
  \item Our work does not cause additional harm to the individuals in our dataset, beyond perpetuating its use in academic literature.
\end{itemize}

Thus, while we are unable to uphold the principle of autonomy to the greatest extent, we believe our research is nonetheless appropriate.

Our analysis of facial recognition systems focuses on racial biases in algorithms, despite evidence that system performance is more closely tied to skin-tone with race as a proxy. However, as our very large reference data sets are racially coded rather than coded by skin tone, a more sophisticated distinguishing analysis was outside our scope.

We restrict ourselves to the law and regulated professions sub-sector of education, centering our work on a limited set of products to more comprehensively cover our chosen area and produce timely research in light of the current social need.

We intentionally refrain from evaluating server side components and functionality of the software packages to avoid the accompanying legal and ethical concerns.  We conducted responsible disclosure to the proctoring suite vendors to make them aware of the vulnerabilities we found along with potential remediations they could take. We believe this mitigates any potential harm disclosing these vulnerabilities may have on exam integrity.  

\section{Related Work}

Several previous studies~\autocite{langenfeld2020internet,turani2020students} have discussed the different threat model remote proctoring solutions face and provided recommendations for security features that could mitigate these new vulnerabilities such as improved authentication measures or 360 degree cameras. \Textcite{slusky2020cybersecurity} extended this by investigating the security feature claims of 20 different exam proctoring suites, noting their strengths and weaknesses against various threat models. \Textcite{teclehaimanot2018ensuring} studied eight John Madison University professors and determined that a significant number of their students seemingly gained an advantage on remotely proctored exams despite the use of a proctoring suite. \Textcite{cohney2020virtual} performed a multidisciplinary analysis of the risks faced by institutions as they rapidly adopt EdTech in light of COVID-19.

A few studies attempted to quantify how a student's perceived stress and performance varies between remote and in person exam environments.  \Textcite{karim2014cheating} studied 582 subjects taking a cognitive test in each environment and found the scores were similar between the two groups but that subjects in the remote setting indicated a higher perceived stress level. \Textcite{teclehaimanot2018ensuring} performed a meta analysis of test scores recorded in proctored and unproctored environments, finding a 0.20 STD variation between the two sets.

\Textcite{teclehaimanot2018ensuring} conducted a survey of eight experts from different universities with previous experience using remote exam proctoring solutions and found the perceived trust of the vendor and the security of the offering to be the primary factors influencing their solution adoption decision. The recent work of~\textcite{balash-proctoring} presented an in depth user-study of student responses to remote proctoring software in light of the pandemic. \Textcite{barrett} discusses the risks of adopting remote proctoring software in terms of bias, test-taker anxiety, and student privacy.

Previous studies have demonstrated biases in the different components of facial recognition systems \autocite{leslie2020understanding,wu2020gender}.  \Textcite{singh2020robustness} extended this by creating targeted attacks to cause false positives by the classifier.  We do not investigate these attacks in the context of Examplify's classifier but anticipate no reason they would not be applicable.  \Textcite{nagpal2019deep} evaluated several different machine learning classifiers using the Multi-PIE and Morph-II datasets and found significant racial biases.  We closely structure our methodology for detecting biases in Examplify's classifier to this work. The National Institute of Standards and Technology (NIST) conducted the Face Recognition Vendor Test (FRVT) to quantify face recognition accuracy at a very large scale \autocite{grother2014face}, providing methodological expertise that guides our work.

Perhaps more importantly, teachers and students have documented their concerns in increasing number since the start of the pandemic. Students have performed their own small scale experiments testing various proctoring suites \autocite{feathers,studenttweet} while teachers have voiced misgivings \autocite{techrev} about use of facial recognition.

\section{Preliminaries}

Computerized exams coupled with the need for remote testing have prompted increased reliance on proctoring software, inclusive of facial recognition classifiers, bringing concerns over privacy, security, racial bias, and fairness to the forefront.

\subsection{Exam Software}
Computerized exam software generally consists of a user interface that displays multiple choice questions with answers and/or text boxes for student answer entry, coupled with a series of cheating detection and deterrence tools tailored to meet the assumed threat model. The general assumption:  students cheat by searching the internet, opening documents/programs on their computers, or consulting a device, person, or printed material during the exam. To this end, exam software generally block access to the internet and non-approved applications on the user's machine; perform audio recordings to detect verbal communication with another person; and run facial recognition to ensure the appropriate individual takes the entire exam and does so without looking away to consult another information source.

Facial recognition is a particularly problematic aspect of exam software given general concerns with the accuracy of such models across different skin-tones and face structures. As a result, individuals from certain groups may be flagged more frequently for potential cheating---a classic case of `disparate impact'.

As a highly regulated industry, law pays particular attention to ethical and professional standards such that when a student cheats within law school or during a licensing exam, there is a high likelihood that the accrediting organization will bar them (potentially permanently) from practice.  We infer that these high standards within the profession affect decisions made regarding exam administration and the choice of what remote proctoring software platform to use, if any.

\subsection{Legal Education}
When we model the threats to law school and bar exams, we take into account the structure of testing in the legal profession and the nature of the exams themselves---which differ substantially from many other fields. Law school course grades depend primarily on a single, tightly curved final exam usually worth at least 85\% of the course grade. Law school grades, and thus exams, are closely tied to job prospects (more-so than many other fields) \autocite{prelaw2019} and bar exam scores are a determining factor in lawyer licensing. These factors, combined with the high-debt burden associated with U.S. legal education, place extreme pressure on students, often providing strong motivation for unscrupulous students to cheat, even by means of paying outside individuals for materials, technical bypasses, or cheating tools. Some law courses utilize multiple choice exams, but more often law exams take the format of a story riddled with legal issues that a student must identify and analyze. Exam answers often consist of several pages of written text so that cheating by simply copying another student's answer would be painfully obvious. Student cheating by attaining/predicting exam content in advance to pre-write answers or using messenger apps during the exam to consult friends comprise the likely, albeit more difficult to detect, form of dishonesty.

COVID-19 has caused bar associations and schools to rapidly adopt an at home remote testing model, forcing administrators to look for additional assurances that cheating attempts will be detected.

\subsection{Software Usage Survey}
\label{sec:survey}
To determine the targets for our technical work, we survey software usage across bar associations and law schools. We identify the remote proctoring software used by the top 180 law schools by scraping public facing websites and making private inquiries to law school administrations.  \Cref{tab:lawSchoolAdoption} depicts our survey results.  We repeat this process for every state bar association's licensing exam and illustrate the results in \Cref{fig:barexamadoption}.  Our data set with the individual school and state bar remote proctoring suite adoption choices is available at \href{https://github.com/WWP22/ProctoringSuiteAdoption}{github.com/WWP22/ProctoringSuiteAdoption}.  Through this process we find that Examplify, ILG Exam360, Exam 4, and Electronic Blue Book comprise the four primary exam software suites used by over 93\% of law schools and 100\% of bar exam associations.  We select these for analysis in the remainder of this work.

\input{tables/lawSchoolAdoption}

\section{Cheating}
\label{sec:security}

The uncontrolled setting and hardware of the remote environment makes for a broad threat model. To evaluate the security provided by these exam proctoring suites we present a reverse engineering methodology and propose three theoretical but realistic adversaries to a remotely proctored law exam. Practical compromises to the security features we identify along with potential attacks from each of these adversaries are discussed.

\subsection{Methodology}

The four exam suites analyzed are not open source, so we reverse engineer the binaries using existing static/dynamic analysis tools. We focus on three questions:  (1) Do the exam suites provide the security guarantees they promise?  (2) What privacy and security is the user required to give up to achieve these guarantees?  and (3) Are the exam integrity checks performed fairly across all examinees?

We isolate suspected critical functions using common reverse engineering methods such as system log inspection and recording of user interface dialogues. We then use traditional disassembly and reverse engineering workflows to manually inspect binary areas of interest.
We employed dynamic analysis to identify and describe functionality that was not apparent through the static analysis, disabling anti-debugging functionality when present in the software.

To evaluate the integrity of exams in transit, we use standard network interposition techniques to evaluate whether the connection to retrieve the exam is over an encryption TLS connection or in plaintext. If the connection is over TLS, we evaluate the program's response to being served:  (1) a valid certificate that has an incorrect common name; and (2) a self signed certificate that is not recognized by any certificate authority but that bears the correct common name. We also attempt to force a downgrade by blocking access to the port before the handshake can occur and then watching for further plaintext retries on a separate port.

For readability, we defer exam fairness methodology to \Cref{sec:fairnessaccuracy} due to its increased depth.

\input{figures/barExamAdoption}

\subsection{Threat Model for Exam Proctoring}
We informally model three adversaries likely to interfere with the fair and secure operation of a remotely proctored law exam. We emphasize that in reality, \textit{many} students may be appropriately modelled as a more sophisticated adversary than their background would suggest by hiring help.  Attack budgets ranging from a few to several thousands of dollars are feasible, given the hundreds-of-thousands of dollars spent on law school.  We define three adversaries:
\parhead{Law Student}
An individual able to adjust basic system and file settings, configure simple hardware devices, and use known passwords but with no experience in reverse engineering software, programming, modifying binary settings, or extracting encryption keys.
\parhead{Law Student with CS Background}
An individual with significant programming experience, familiarity with basic system administration, and the knowledge to extract keys from the binary but with no ability to modify any portion of the binary and without extensive experience reverse engineering software.
\parhead{Experienced Reverse Engineer}
An individual with all the prior capabilities with additional experience analyzing and rebuilding binaries, disassembling software, and applying patches.  Familiarity with advanced system administration and the ability to:  adjust any setting, configure hardware devices, modify drivers, use custom encryption tools, and extract encryption keys from and modify the control flow of the binary is assumed.  While we expect the number of students with this background is low, such individuals may sell their services or cracked versions of software.

\subsection{Reverse Engineering Findings}
In this subsection we detail our findings regarding how proctoring suites operate and discuss each suite's security features and implementation.

Exam proctoring suites use a few distinct components to ensure exam integrity is maintained: \textbf{computer monitoring}, \textbf{exam content protection}, and \textbf{identity verification and authentication}. We categorize our findings based on these components for the following four proctoring suites:  Examplify, Exam 4, Electronic Blue Book, and ILG Exam360.

\subsubsection{Computer Monitoring}

The computer monitoring components in an exam suite aim to prevent a student from accessing unauthorized resources during the exam and may even restrict the student from beginning the exam if certain parameters are not met. A summary of the features used by each suite is provided in \Cref{appendix:features}.  The features we identify as comprising the computer monitoring component of exam integrity are detailed with each exam suite's feature implementation outlined:

\parhead{Virtual Machine Detection} A student running the proctoring software inside a virtual machine environment can easily exit the environment with a hotkey and completely evade any monitoring the exam proctoring suite hoped to provide. To prevent this, most suites feature virtual machine detection to detect and prevent attempts to run the software inside a virtualized container. The most common implementation of this is a simple check of the computer's CPU manufacturer to see if the string matches known virtual machine software vendors. An additional check of CPU temperature for constant values or known presets can be run since a virtual CPU does not often pass through the real value from the actual CPU.

All of the exam suites we examined implement a virtual machine check by comparing the CPU vendor field against a list of known virtual machine vendors.  Examplify extends this by retrieving the CPU temperature and flagging a device if it reports a CPU temperature of 100C as this is the typical default value virtual machine vendors use.  Electronic Blue Book also checks the computer, hard drive, network adapter, and bios vendor information to see if any fields contain the string `virtual'. If a virtual machine is detected, they log the attempt and prompt the user to run the software outside of a virtual machine. \Cref{appendix:vmblock} provides a summary of the popular virtual machine software blocked by each proctoring suite.

\parhead{Virtual Webcam/Microphone Detection} Virtual webcam and microphone programs exist that allow a user to generate a virtual device which either returns data from a remote endpoint or from a file.  Adversaries could bypass exam suite identity verification by returning video of themselves in another location while someone takes the exam for them or by returning a prerecorded video of them taking the exam.  Exam proctoring suites attempt to mitigate this threat by checking the device vendor and bus location against a list of flagged vendors.  If one of these blocked devices is detected, the software will either flag the exam for further review or prevent the student from beginning to take it.

Examplify and ILG Exam360 detect virtual webcams and microphones by retrieving the operating system's device list and comparing it to known virtual device vendors. Exam4 and Electronic Blue Book do not use the computer's webcam or microphone so they don't implement a check. \Cref{appendix:vwmblock} provides an overview of the popular virtual webcam/microphone vendors on the blocked list.

\parhead{Clipboard Management} Students copying pre-written text into an exam is a major concern, especially in the field of law where exam essay answers may require lengthy summaries of law and/or analysis that can be prepared before the exam. Exam proctoring suites attempt to prevent this by either clearing the clipboard before the exam or logging its contents for later analysis.  During the exam the clipboard generally can be used inside the proctoring suite but copying from outside apps is prohibited or logged using a similar method.

The exam proctoring suites implement clipboard protection by calling the system clear function.  The content is not captured before the clear operation by any of the suites.  Exam4 and ILG Exam360 implement a custom restricted clipboard for use inside the test environment which limits what can be copied.

\parhead{Screenshot Capture} Exam proctoring suites may offer screenshots of the student's screen during the exam to allow a proctor to retroactively review the exam session to determine if unauthorized resources were accessed on the computer. These screenshots are normally captured using a highly privileged system level service which leads to potential privacy issues when an exam is not in progress.

ILG Exam360 is the only exam suite we analyzed that provides screenshot captures of the student's computer during an exam.  The screenshot is captured by calling their highly privileged system service using a Javascript call which uploads the screenshot to Exam360.

\parhead{Process/Application Restrictions} Process restrictions are normally used to limit what applications a student can access during an exam. These are generally implemented using a process allow list that contains processes specifically allowed by the exam in addition to critical processes the operating system needs to maintain the computer's function. A weaker implementation involves using process block lists that prevent certain processes such as web browser activation from being started.  Both approaches are implemented using the exam proctoring suite's highly privileged system service which starts a watchdog service that forcibly kills unauthorized processes.

Examplify and ILG Exam360 compare the processes currently running on the system to a list of processes they and the exam itself allow, continuously monitoring for and forcibly killing any processes that are running but not included on the list.  Exam4 and Electronic Blue Book have a list of disallowed services which they kill upon exam commencement.

\parhead{Network Interception} Closed book exams require the software to  limit a student's ability to search for information on the internet. Approaches to block internet access that we found included:   dropping internet traffic, inserting an active man in the middle to capture traffic, or redirecting the traffic to the vendor's servers.  The simplest approach is dropping the traffic using a routing rule.

Examplify restricts network traffic by inserting a null route into the default operating system routing table. Exam4 disables the network adapter during the examination. ILG Exam360 inserts a null DNS entry into the network configuration causing domain name resolution failure. Electronic Blue Book does not implement any network restrictions other than blocking access to common websites through their process block list. None of the implementations inspected capture browser traffic or redirect it to a site controlled by the suite vendor.

\subsubsection{Exam Content Protection}
Exams are often downloaded to student computers before the actual exam begins making security during transit between proctoring suite vendor servers and students paramount since traffic can be easily intercepted using off the shelf solutions.  Exam protection during the download and while it sits on the student's computer is vital to prevent early or unauthorized access. We detail below our findings for each of the exam suites for both encryption at rest and in transit.

\parhead{Encryption In Transit} Examplify and ILG Exam360 use transport layer security (TLS) for all of their connections. The certificate chain, expiration date, and common name are correctly verified which mitigates active man in the middle attacks. The connection is never downgraded to plaintext HTTP even if the software is unable to successfully complete the handshake.  Examplify includes their own certificate store inside the software to prevent potentially using a modified system certificate store. Exam4 and Electronic Blue Book allow the individual institution to select their transport layer security settings. Electronic Blue Book allows each institution to configure whether to use TLS or not, and the configurations are available publicly. We found several that failed to enable TLS. Exam4 similarly features per-school configuration, but the configurations are not public. However, the school from whom we obtained the Exam4 binary did not have TLS enabled.

\parhead{Encryption At Rest} Examplify, ILG Exam360, and Exam4 use AES-256 for encrypting exams at rest on student computers. ILG Exam360 and Exam4 use SHA1 to derive the AES key from a password.  Examplify uses 10,000 iterations of the PBKDF2 function to derive the exam password.  The exam manifest which contains the main exam password salt, exam information, allowed resources, and monitoring settings is encrypted separately using a static key stored in the Examplify binary. Electronic Blue Book allows the institution to choose between 3DES and AES for exam encryption and uses 1,000 iterations of SHA1 by default for password derivation, but the institution can configure the iteration count. The password salt is statically stored in the Electronic Blue Book binary and is set to `Kosher'.

\subsubsection{Identity Verification and Authentication}
Exam suites all implement some form of user authentication to ensure the test taker matches the individual to be assessed.

\parhead{Logins}Exam4, ILG Exam360, and Electronic Blue Book implement standard single factor logins. Examplify implements a similar single factor login.  OAuth is not supported by any of these solutions so institutions cannot easily add more extensive identity verification measures such as two factor verification.
\parhead{General Interaction Fingerprinting} General interaction fingerprinting analyzes the pattern of a student's key strokes and mouse movements against the class average for anomalies and if present, flags the exam for human proctor review. This poses the risk of potentially unfairly flagging students with disabilities or those who legitimately deviate from the class average pattern. While none of the suites we analyzed use this, it is used by Proctorio, a common proctoring suite outside our scope.
\parhead{Facial Recognition}As an analog to students showing ID upon entering an exam hall, some exam proctoring suites employ facial recognition for identity verification to counter another person taking the exam for a student. The distance of the facial feature vectors of a student's image are compared against those of the student's trusted reference image and if below a certain threshold, the student is considered verified.

ILG Exam360 offers facial recognition but also employs a remote human verification method before exam initiation utilizing a webcam to connect a student with a human proctor who conducts the final verification. In the end, the process resembles that of an exam hall.  Examplify's verification implementation relies on an automated facial recognition classifier.  Our research quantifies bias introduced by Examplify's process which we detail in \Cref{sec:examplifyfairness}.  Exam4 and Electronic Blue Book do not offer facial recognition.

\subsection{Impact Analysis}
To evaluate the impact remote exam proctoring has on exam integrity, we must determine whether its security features are effective against expected adversaries; whether privacy concerns are generated by these features; and whether these features introduce bias into the exam taking process.

\subsubsection{Security Feature Vulnerabilities}
Proctoring suite security features have various vulnerabilities depending on the expertise of the adversary. We validate the attacks referenced in this section against the proctoring suite in a simulated test environment to ensure they can actually be conducted.

\parhead{Virtual Machines} Virtual machine software allows a guest operating system to be run inside the primary environment, bypassing any monitoring the proctoring suite could hope to achieve. The CPU vendor check implemented by all of the exam suites can be easily bypassed using common virtual machine software since this field is generally configurable. The CPU temperature check Examplify conducts is defeated by configuring the virtual machine to pass through the CPU temperature of the actual computer's CPU vendor or a random string not on the block list the exam proctoring suites are using. A law student with a CS background, capable of installing a virtual machine and configuring the CPU vendor, could readily complete this attack.
\parhead{Virtual Webcams/Microphones} Virtual webcam and microphone devices can allow adversaries to take exams outside an appropriate physical context by replaying a prerecorded file or piping a connection to a separate device. The only mitigation employed to prevent virtual devices is a driver list search which looks for known virtual device vendors along with the running process restrictions. A student can evade detection easily by installing a virtual driver not on the known vendor list or more skillfully by renaming the driver of a blocked virtual device since the signing key of the driver is not checked by the exam proctoring suite. All these attacks are well within the capabilities of a student with a CS background.  A student able to create a custom virtual device that masqueraded as a legitimate driver could redistribute this to other students.
\parhead{Clipboard Protection} Exam proctoring suites protect against clipboard content being carried into the exam by calling the operating system's clear function before the exam begins. This does not preclude the use of an external hardware connected clipboard such as a KVM or a built in keyboard macro which allows note storage. The majority of these hardware connected devices simply present as standard hardware interface devices which don't require any additional drivers. Exam proctoring suites could attempt to protect against clipping and pasting clipboard content by fingerprinting the input rate of a student's keystrokes. Purchasing a hardware device capable of maintaining an external clipboard is an attack most law students could perform. We do not investigate the Mac version of the exam proctoring suites but colloquial evidence suggests that the iCloud clipboard sharing might bypass these protections by loading information from the phone's clipboard into the Mac keyboard through the service during the exam.
\parhead{Process Restrictions}  Process restrictions are implemented either with a list of allowed processes or with a list of disallowed processes that are killed upon starting the exam. To subvert either restriction, a student with CS background could recompile a piece of open source software to report a different process name. As an example, Chromium could be recompiled to report as `explorer.exe' which is allowed by every testing suite we looked at since it is a critical user interface component for Windows based systems. For suites using block lists, most law students would be capable of finding a process not present on the disallowed list through trial and error.

\subsubsection{Student Privacy Concerns}
Continuing to evaluate the impact remote exam proctoring software use has on exam integrity, two major privacy questions arise:  (1) Is the user appropriately informed of the information being collected upon engaging with the remote exam software?  and (2) Does the potential for pervasive monitoring after the student is no longer actively taking an exam exist?  To this end, we develop an analysis tool to assist other researchers in identifying remote exam software privacy issues.

\parhead{Informed Consent}
\input{tables/datasetSummary}
An examinee cannot provide meaningful consent to the activities performed by the exam proctoring software if they are not informed of the specific data being collected or the surveillance mechanisms performed. Examinees are prohibited from reverse engineering the software to discover such information and attempts to glean this information by reading privacy policies, end user license agreements, or similar documents will be met with vague and sometimes conflicting verbiage. As an example, ExamSoft's privacy policy notes, ``in order to secure the exam taker’s device, ExamSoft must access and, in some instances, modify device system files.'' This broad statement provides no substantive limitation on what the Examplify software may do. Other privacy policies contain conflicting statements about the software's activities. For example, the Extegrity Exam4 privacy policy states, ``Exam4 does not access any data on the laptop other than the data created by the act of taking an exam'' and ``Exam4 monitors activity the student engages in while taking an exam and creates an encrypted log report for evidentiary purposes.'' It is not possible for Exam4 to monitor activity the student engages in if it does not access any data other than that created by the act of taking an exam. These types of statements thwart a student's  ability to meaningfully consent.
Furthermore, even if an examinee was fully aware of the software privacy implications, meaningful consent is still lacking given the examinee's lack of meaningful alternatives.  Faced with accepting and using the software as-is or refraining from taking the exam, most law students would opt for the former as the latter coincides with an inability to become a licensed lawyer. Such a choice is not a choice. While legal agreements offer a degree of liability cover for the exam software companies, they fail to meet conventional ethical standards for consent \cite{appelbaum1987informed}.
\parhead{Post Exam Monitoring} Examplify installs a highly privileged system service that is constantly running on the computer even if Examplify isn't open. Currently running applications on a user's computer are logged to a debugging file that is uploaded periodically once the Examplify application is open. The service also regularly reaches out to the Examplify server to check for and install updates for the service or the binary. Exam4 and ILG Exam360 also implement a system service but stop it when the exam is terminated gracefully. Electronic Blue Book directly hooks into the Windows system service with their binary to provide their monitoring features, guaranteeing no additional background monitoring is being performed once the binary is closed.

\subsection{Automating Impact Analysis}

We created an analysis tool based on the RADARE2 and Ghirda frameworks \autocite{cheng2016binary} to simplify the reverse engineering process for researchers who want to quickly analyze the privacy impact of other exam proctoring solutions. We release the tool publicly at \href{https://github.com/WWP22/ProctoringSuiteAnalysisTool}{github.com/WWP22/ProctoringSuiteAnalysisTool}. The approach we use in this paper using traditional tools like IDA work well for in depth studies, but they are not well suited for providing a quick summary of an exam proctoring suite's privacy impact. Our tool requires a user to simply run the Python script on the binary they want to analyze and a high level overview of the application will be provided.

\subsubsection{Design}

The analysis tool first loads all of the shared object files the binary uses and then performs auto analysis using RADARE2's built in analysis suite.  This attempts to locate the segments and functions to generate a control flow graph.  From this control flow graph, we extract cross references to lines in any part of the code which allows us to more easily establish where certain data elements are being used.

We are able to detect privacy sensitive calls such as calls to a microphone, webcam, or video driver by fingerprinting common vendor and system library names. We also attempt to extract information about the security features the exam proctoring suite implements including whether it detects virtual machines, uses a secure connection to reach the back-end server, and whether it encrypts on disk content. If on disk encryption is found, we display the cipher suite being used then attempt to extract the encryption key and initialization vector by searching for keys of the correct bit length in a user definable window around any data references found in the encryption function. For a more complete analysis, the tool can be run with the live memory option which initializes the binary, attaches GDB, and runs to a user defined breakpoint then performs the analysis.  This allows a more complete analysis of libraries and code segments which are stored encrypted at rest or loaded from a remote endpoint. A user can view a summary of the binary's security and privacy properties or opt for a more in depth analysis featuring control flow graphs and decompilations with Ghirda of code segments the tool finds relevant.
\input{tables/fnmrFmrSummary}

\subsubsection{Evaluation}

We evaluate the automated analysis tool on the four suites to determine whether it accurately identifies relevant security and privacy information. The tool is evaluated without using the live memory analysis option since we believe this would be the most commonly run configuration of the tool due to the relatively large performance and memory overhead of searching the entire live memory space multiple times on consumer hardware.

We find the tool able to correctly identify camera and microphone usage in the all cases aside from a false positive triggered when analyzing Electronic Bluebook.  This false positive is due to the inclusion of a large English dictionary in the binary which incorrectly triggers one of the vendor searches we run. The relevant function control flow graph and decompilation is presented to the user which would allow the user to trivially identify it as irrelevant and subsequently disregard.

The tool performs similarly well with virtual machine detection, insecure connections, and on disk encryption with results mirroring the results we obtained in our manual analysis. The encryption key Examplify uses is successfully extracted and presented to the user along with a few false positives. We write a simple Python script to test the encryption keys and initialization vectors the tool outputs and are able to successfully decrypt the on disk libraries for Examplify in under one minute.

\section{Automated Identity Verification}
\footnotetext[2]{VMER uses more specific race annotations than the current federal government standards for collecting and presenting data on race.  We group these more specific annotations as follows using the federal standards and the NIST FRVT as a guideline: Caucasian Latin$\rightarrow$  White; African American$\rightarrow$  Black; East Asian$\rightarrow$  Asian; and Asian Indian$\rightarrow$ Indian\label{vemr-labels}}
\label{sec:fairnessaccuracy}

Traditionally, human verification through simple recognition of a student and identification card checks have been used to ensure exam integrity.  While these checks could still be conducted in the remote setting these exam software suites operate in, there is a large incentive for exam proctoring vendors to move to a fully automated model using an automated facial recognition system due to reduced staffing cost and an increased capacity for the number of students that can be verified. When facial recognition classifiers are used to determine who can take an exam or whether a student is flagged for cheating, it is critical to the fairness of the examination to insure that the system is accurate and fair.

We evaluate the overall accuracy of current state of the art facial recognition systems along with the facial recognition classifier used by Examplify, `face-api.js', over several datasets we believe accurately reflect the real world images these systems would be operating on.  The classifiers we select for comparison against `face-api.js' are based on their accuracy using Labeled Faces in the Wild (LFW) \autocite{serengil2020lightface}, the current dataset used most prominently in the literature for benchmarking the fairness of facial recognition algorithms \autocite{kemelmacher2016megaface}. This results in a selection containing FaceNet \autocite{schroff2015facenet}, VGG-Face \autocite{parkhi2015deep},  OpenFace \autocite{amos2016openface}, and ArcFace \autocite{deng2019retinaface, guo2018stacked, deng2018menpo, deng2018arcface}.

Given that the expertise of remote proctoring firms is outside AI/ML and that in the current business environment such firms are unlikely to develop and train their own models (an assumption borne out by our analysis of the leading market product), it is reasonable to select pre-trained, off-the-shelf models for comparison. Finally, we conduct an analysis of the error rates based on the subject's race which allows us to look at the fairness of the classifier across subjects from different races. We release our image selections, data processing code, and classifier performance results at \href{https://github.com/WWP22/AIVerify}{github.com/WWP22/AIVerify}.

\subsection{Accuracy}

\label{sec:accuracy}

Using the knowledge gleaned from our analysis of Examplify's identity verification system, we separate the facial recognition steps an exam proctoring suite would need to perform into two steps: the initial verification against a student's identification photo to bootstrap their identity; and the continuous verification that insures the student who verified their identity initially continues to take the entire exam.

\parhead{Dataset Selection} We select MORPH-II \autocite{morph}, Multi-PIE \autocite{gross2010multi}, LFWA+ \autocite{liu2015faceattributes} and VGGFace2 \autocite{cao2018vggface2} as our datasets to evaluate the performance of the facial recognition classifier using images that are very similar to the ones likely to be encountered during real world use. A full description of the datasets can be found in \Cref{tab:datasets}.

\parhead{Metric Selection} We focus on the false non match rate (FNMR) and the false match rate (FMR) of the facial recognition systems for our evaluation. We select these metrics as they allow us to determine the rate at which the facial recognition system would fail to verify a student (FNMR) and at what rate the system would falsely verify a different person as the student (FMR). These metrics are also used in other key studies on facial recognition accuracy such as the NIST FRVT \autocite{grother2014face} which allows our results to be easily comparable.

\subsubsection{Initial Verification}
\label{sec:initialver}
Initial identity verification in a remote exam setting relies on comparing a known image of the subject (likely a school ID or driver's license) to a capture of the subject trying to take the exam. This can present background, lighting, and age progression challenges for the facial recognition classifier.

\parhead{Image Selection} We select images from each of the above datasets to create sub-sampled datasets for our FNMR and FMR evaluation. We create our FNMR datasets by selecting the earliest capture of each subject in the dataset as our reference image. We use up to 50 images from subsequent captures of the same subject to compare against.  The extended length of time between subsequent captures in the Morph-II, LFWA+, and VGGFace2 datasets make them useful for the initial identity verification comparison since institutions commonly use the student's driver's license which has an average photo refresh time of 5.7 years \autocite{tefft2014driver} as the reference image. For the Multi-PIE dataset we select subjects who attended multiple sessions, using each subject's first attended session as the reference image and up to three of their subsequent session images as comparison images. Use of these Multi-PIE captures provides insight on whether more recent reference images significantly improve the performance of the facial recognition classifier  We create the datasets we use for evaluating the FMR by selecting the earliest subject capture as the reference image and 50 earliest captures of other subjects as comparison images. \Cref{appendix:ivimageselect} summarizes these sub-sampled datasets.

\parhead{Classifier Performance} High FNMRs occur across all of the datasets evaluated for both the SOTA classifiers and `face-api.js'.  When run on datasets that are in-the-wild versus those with more constrained capture conditions, significantly elevated FNMRs occur across all of the classifiers except VGG-Face.  A significant difference in the FNMRs between the Morph-II and Multi-PIE datasets was not noted which suggests that the time between the reference image and the comparison image is not a major factor in classifier performance.  Overall, we see an inverse relationship between the FNMR and the FMR such that an algorithm less likely to fail at identifying a correct student would be more likely to verify an imposter (hired test-taker). We summarize these findings in \Cref{tab:fnmrfmrtable}.

\parhead{Analysis} These high FNMRs make any of the classifiers we evaluated inappropriate for use in an automated facial recognition setting due to the extreme cost to a student unable to be verified. Exam proctoring vendors can attempt to reduce the error rate by selecting classifiers with the lowest FNMR, however, these classifiers all present much higher FMRs suggesting they provide little more than security theatre. Using more recent trusted reference images had minimal impact on the FNMR when compared to using images captured over longer time periods  minimizing an institution's ability to mitigate the high FNMR by collecting more recent reference images.

\subsubsection{Continuous Verification}
\input{tables/fnmrFacialRotation}

Continuous identity verification works to ensure the initially verified student takes the entire exam and is not replaced by another person.  The student's recent reference image verified during initial verification is compared to subsequent captures taken silently at random intervals during the exam. These unprompted captures create challenges for the facial recognition classifier since the student's facial tilt, pose, and lighting may all vary significantly from the prompted reference image taken in a controlled fashion.

\parhead{Image Selection} We create subsets from the Multi-PIE dataset for evaluating FNMR based on varying facial rotations, facial expressions, and lighting. \Cref{appendix:cvimageselect} describes each dataset.

\parhead{Classifier Performance} The performance between the SOTA classifiers and `face-api.js' is roughly equivalent up to a 30 degree head or facial rotation but diverges thereafter with `face-api.js' maintaining a relatively low FNMR up to 60 degrees.  The SOTA classifiers begin failing to verify the majority of subjects once the rotation reaches 60 degrees while `face-api.js' still verifies over a 75 degree rotation. \Cref{fig:facialrotation} shows the relationship between facial rotation and classifier performance. There is minimal variation from the average FNMR discussed in \Cref{sec:initialver} when the subject changes their facial expression.  Classifier performance falls off sharply as the lighting conditions differ more significantly from the reference image. `Face-api.js' significantly underperforms VGG-Face when evaluating images taken under varying lighting conditions but outperforms all of the other SOTA classifiers. \Cref{appendix:cvperf} provides a full description of the FNMR performance on these datasets.

\input{tables/fnmrRace}

\parhead{Analysis} The classifiers exhibit significantly increased FNMRs once they are tasked with comparing images in challenging conditions such as varied lighting or head/facial rotation.  Students taking an exam in good faith over long exam periods have images that reasonably exhibit lighting and head/face position variation especially since captures are taken without notice.  The FNMR exhibited even in the less extreme conditions suggest that a significant number of students would fail to be verified correctly.  A majority of students would fail to be validated in the more extreme conditions we tested.  The variance in lighting seen was on the more extreme side but not outside the realm of real world exam environments where exams can start in daylight and end after dark.

\subsection{Fairness}

When forms of identity verification are used as part of examination procedure, it is critical to the fairness of the examination that every effort to minimize bias is made.

\parhead{Dataset Selection} We select MORPH-II, LFWA+ and VGGFace2 as they either provide labels of a subject's race in the original dataset or have labels available through additional studies such as VGG-Face2 Mivia Ethnicity Recognition (VMER)
\autocite{Greco_MVA2020}.  Multi-PIE is not included in this section due to the limited number of subjects in some of the races we analyze.

\parhead{Metric Selection} The FNMR metric, which potentially flags a student for cheating, is utilized in this section since it reflects the most negative impact on the student.

\parhead{Classifier Performance} On the Morph-II dataset all classifiers except OpenFace flag `White' subjects at a slightly higher rate than `Black' subjects. Significant outliers in the Facenet and Arcface classifiers for `Asian' subjects are evident. Across races, `Asian' subjects are flagged at significantly higher rates in all classifiers except FaceAPI.  The LFWA+ dataset shows a much wider distribution of FNMRs within a racial grouping across classifiers and  higher FNMR values for `Black' subjects versus `White' subjects occur for all of the classifiers except VGG-Face and Facenet.  Compared to other races, `Asian' subjects have lower FNMR values across all classifiers run on LFWA+ except VGG-Face.  Much like classifier performance on the LFWA+ dataset, all classifiers run on VGGFace2 except VGG-Face yield higher FNMR values for subjects labelled as `Black' versus `White'.  Additionally, `Asian' and `Indian' subjects also fair better than `Black' subjects on all classifiers except VGG-Face on the VGGFace2 dataset.   While lower FNMRs are seen on `Asian' subjects compared to `White' subjects with LFWA+, elevated FNMRs occur on the `Asian' group compared to the `White' group using the VGGFace2 dataset except with the Facenet and VGG-Face classifiers. `Indian' subjects have the lowest FNMRs across all races using all classifiers except VGG-Face on the VGGFace2 dataset. A summary of these results can be see in \Cref{tab:racefnmr}.

\parhead{Analysis} We see significant variations in the performance of the different classifiers depending on the race of the subject being analyzed. In the large VGGFace2 and LFWA+ datasets, we see significantly higher FNMRs for subjects labelled as `Black' versus `White'.  Similar disadvantage for subjects tagged as `Asian' compared to those tagged as `White' occurs on the VGGFace2 dataset.  This demonstrates the propensity for minority groups to be flagged at higher rates than other groups depending on the capture conditions. Additionally, we see major variance in FNMR performance between subjects of different races based on the dataset, demonstrating the difficulty of correcting for this in the software using the classifier.

\subsection{Evaluating Real World Implementation}

\label{sec:examplifyfairness}

Examplify uses the `face-api.js' classifier for facial recognition with a pretrained publicly available model inside the exam proctoring suite. We see in \Cref{sec:accuracy} that `face-api.js' was quite inaccurate overall with an average FNMR of 11.47\% when evaluated on datasets that accurately reflect images it would be running on in a testing environment. In Examplify's implementation, the initial verification step acts as a gate keeper for the student to be able to take the exam.  The high average FNMR suggests that this would disadvantage numerous students attempting to take their exam. We see similarly concerning performance on the continuous verification task which runs in the background during an exam if enabled by the institution. Realistic variations in the capture conditions such as head/face rotation or lighting changes could cause the image to fail to be verified, flagging the student's exam for human review. Depending on how this is presented to the proctor for human review, this may unfairly bias the proctor against the student with a presumption of guilt.

We see significant variations in the performance of the `face-api.js' classifiers depending on the race of the subject with some minority groups being disadvantaged in LFWA+ and VGGFace2 datasets. This suggests that the automated facial recognition system may be unfairly disadvantaging certain students based on their race by not allowing them to take their exam or by presenting their exam to a human proctor for review at a higher rate than other non-minority students in certain cases.  Given the variability and bias in the facial recognition steps, we believe a human based verification model is a significantly fairer approach to insuring exam integrity.

If an automated classifier is to be used, we recommend training models on a dataset that contains a balanced sampling of subjects from different races versus using pretrained default models. We also recommend evaluating the performance of the classifies on datasets that realistically represent the use case of the system. Classifier performance cannot be accurately assessed just using an overall performance metric without looking at that performance metric across subjects from different races to determine whether the classifier is acting in a fair manner.

\section{Discussion}

\parhead{Impact on Marginalized Groups}
Minorities and other marginalized groups are traditionally underrepresented in the legal profession. Exam software with the built-in skin-tone biases creates an invisible barrier. This occurs both in the intuitive case where minorities are flagged for cheating at higher rates, but also where they are flagged at substantially \textit{lower} rates. Opponents of affirmative action have erroneously argued that minorities are not able to cope with the rigor of law school. By using software that flags minority students substantially less, such opponents will continue to cast doubt on the validity and competence of minority students. Thus, substantial bias in either direction can perpetuate systemic racism.  This may further impair the willingness of minorities to enter the legal profession.
Law schools, bar associations, and other educational/licensing institutions must investigate and commission research into inherent bias in the exam software they utilize to prevent discrimination in exam administration.

\parhead{Fundamental Challenges of Remote Examination}
Remote exam proctoring suites suffer from fundamental limitations in the threat model they can protect against since they run on untrustworthy student hardware versus exam hall hardware. The student has full administrative access and will always be able to bypass security features and monitoring scheme protections given enough time. Exam proctoring suite vendors can attempt to increase the time and skill level necessary to compromise the exam by adding complexity to the process through obfuscation and active anti-debugging measures.

In order to create a truly secure remote exam proctoring suite, a vendor needs to establish a trusted environment on the device that restricts the student's ability to extract or modify part of the exam suite.  Intel SGX and other similar trusted execution environments could potentially be employed for this, however, it may introduce significant usability and availability concerns due to a much more complex bootstrapping procedure. Furthermore, Intel has recently announced the deprecation of SGX in consumer chips---as a result of ongoing security issues with the technology~\autocite{inteldatasheet}. 

\parhead{Risks of Privileged Software}
All the software we evaluated require privileged system access. Operating systems increasingly restrict such access as it is a common source of malfeasance. Buggy but well-intentioned code given such access substantially broadens the attack surface of the OS and serves as a glaring target. Experts urge against granting such access even to third-party antivirus software \autocite{ars-antivirus}! Compounding the problem, students are likely to be unaware that privileged system services from the proctoring packages do not uninstall automatically and persist after the exam is over \autocite{balash-proctoring}, putting them at long term risk.

\smallskip

\parhead{Privacy, Surveillance, and Ethical Concerns}
Attempting to meet their design goals, platforms engage in sweeping surveillance.  Keylogging, screen captures, drive access, process monitoring, and A/V feeds give their software access to personal data stored on the device. Outside of this context, these features appear only in malware, highlighting the unusual capabilities of these software. Some binaries also include anti-debugging mechanisms in their code, further limiting the ability of student advocates to assess the security and safety of the software.

The context in which students are required to install proctoring software mitigates their ability to meaningfully consent to the substantial impositions on their privacy and security. The veneer of legitimacy of the platform conveyed by the institutional backing of the school or testing company compounds this. This dynamic is partially captured in the results of \Textcite{balash-proctoring} who find that trust in institutions substantially reduces the extent to which students are willing to object to remote proctoring.  Even if they did, students are not provided with a reasonable alternative.

\parhead{Recommendations} The strongest recommendation we offer is that where allowable, educators should design assessments that minimize the possible advantage to be gained by cheating. Project work, open book essays, or other similar modes with unrestricted access to resources feature fewer opportunities to gain unfair advantage.

Where re-imagining assessment is not possible, students should be offered a \textit{meaningful} chance to opt-out of digital testing on their own hardware and should be given the choice of either using provided hardware or paper-copy and live proctoring. Furthermore, schools put substantial efforts into helping students install proctoring software and should match this with equal efforts to help them uninstall the software while advising them of the risks of retaining it.

Given the substantial fairness concerns with facial recognition systems, our primary recommendation is to avoid using them. Where infeasible, it is vital that human review remain a central component and that such review be conducted by multiple diverse individuals to reduce human biases as well.  Lastly, until such time as more advanced facial recognition technology is developed, if these current classifiers are going to see continued use, candid conversation regarding generalized differences in facial features between racial groups needs to be addressed by programmers through dialogue with racially diverse focus groups and accounted for in calibration settings in an attempt to reduce the incidence of false identification.

\printbibliography

\newpage
\onecolumn

\appendix

\section{Proctoring Suite Feature Summary}
\label{appendix:features}

We provide a summary of the various security and privacy features each exam proctoring suite we analyzed currently offers.  We categorize the features based on whether they have a direct relation to a user's privacy.  If a feature wasn't implemented by the exam proctoring suite, we mark the feature as not implemented (N/I) for that suite.

\subsection{Security}

\begin{center}
  \input{tables/securityFeatureComparison}
\end{center}

This table provides an overview of the security features provided by Examplify, Exam4, Electronic Blue Book, and ILG Exam360. All of the exam suites use strong encryption algorithms to protect exam content at rest.  Only two exam suites, Examplify and ILG Exam360, use HTTPS for their encryption in transit while the other exam suites rely on the content being protected using the same keys they protect the exam at rest with. Virtual machines are blocked by all of the exam suites using block lists which have common vendors and virtual machine properties listed in them.  Virtual device detection is only implemented by Examplify and ILG Exam360.  Since Exam4 and Electronic Blue Book do not use the microphone or camera, they do not implement any virtual device detection. Clipboard management is implemented by all of the exam proctoring suites by either implementing a custom restricted clipboard or by using the system clipboard with a function call to clear it before the exam begins.  ILG Exam360 was the only proctoring suite that implemented a form of screenshot capture to see what was on the student's screen during the exam.  The processes a student can run are restricted on all of the exam suites either through an allow list which allows certain processes to be run by the student or a block list which specifically prevents certain processes from being run by the student.  Network access restrictions are offered in Examplify, Exam4, and ILG Exam360 while Electronic Blue Book does not offer a way to restrict internet access aside from the process restrictions.

\subsection{Privacy}

\begin{center}
  \input{tables/privacyFeatureComparison}
\end{center}

This table provides an overview of the privacy related features in Examplify, Exam4, Electronic Blue Book, and ILG Exam360. Initial identity verification is only offered by Examplify and ILG Exam360.  Exam4 and Electronic Blue Book do not offer a way to identify a student using any biometrics. Similarly, continuous identity verification is only offered by Examplify and ILG Exam360. For both identity verification steps, Examplify uses an automated method while ILG Exam360 uses a human to complete the verification step. All of the proctoring suites implement some form of system service to allow a significant portion of their functionality.  Examplify's monitoring service was the only service we found to continue running after the application was closed.  Device identifiers are sent by Examplify and ILG Exam360 to a centralized server while Exam4 and Electronic Blue Book do not collect any device metadata.

\section{Virtual Machine Block List}
\label{appendix:vmblock}
\begin{center}
  \input{tables/virtualMachineDetection}
\end{center}

This table provides a summary of popular virtual machine software that is blocked by the exam proctoring suites we evaluated. VirtualBox and VMWare Workstation are blocked by all of the proctoring suites.  VMWare Fusion which is only available on Mac was blocked solely by Examplify even though ILG Exam 360 and Exam4 offer Mac versions.  Parallels was detected by Examplify, Exam4 and ILG Exam360 but not by Electronic Blue Book.  Hyper-V and QEMU were only detected by Examplify.

\section{Virtual Webcam/Microphone Block List}
\label{appendix:vwmblock}

\begin{center}
  \input{tables/virtualDeviceDetection}
\end{center}

This table provides an overview of the popular virtual webcam and microphones that were detected or blocked by the Examplify or ILG Exam360 proctoring suites.  Exam4 and Electronic Blue Book do not use the webcam or microphone during exams so they did not implement specific security features to target virtual webcams or microphones.  ManyCam, YouCam, and MyCam are blocked by both Examplify and ILG Exam360 while OBS Studio is only detected by ILG Exam360.  Logitech Capture is undetected by all of the proctoring suites.

\section{Initial Verification Image Selection}
\label{appendix:ivimageselect}
\begin{center}
  \input{tables/initialVerificationDatasetSummary}
\end{center}

This table provides a summary of the datasets we use for our evaluation of the facial recognition classifier's performance in an initial verification setting. The number of subjects for the FNMR datasets is lower than the FMR dataset since we require subsequent images of the same subject so if a subject only has a single image, they would be excluded. We select the first image of the subject in the dataset as the reference image then compare against up to 50 subsequent images of the same subject for the FNMR datasets.  We compare the subject against 50 other subjects at random for the FMR datasets.

\section{Continuous Verification Image Selection}
\label{appendix:cvimageselect}
\begin{center}
  \input{tables/continuousVerificationDatasetSummary}
\end{center}

This table summarizes the datasets we use for our evaluation of the facial recognition classifier's performance in a continuous verification setting. The `Session' dataset uses the session image of a subject as the reference image and images from all subsequent sessions as comparison images.  We use five different lighting conditions to increase the number of samples with each reference image being compared against a comparison image using the same lighting condition. The number of subjects for the Multi-PIE Session dataset is lower than the overall number of subjects in the Multi-PIE dataset since subjects were only included in this dataset if they attended multiple sessions and only a subset of the total subjects attended multiple session. The `Facial Rotation' dataset uses an image of the subject facing directly into the camera as the reference image with each comparison image being another image of the subject at a different rotation away from being centered and facing towards the camera. The `Lighting' dataset uses a well lit image of the subject as the reference image then four comparison images with different degrees of lighting which cause partial feature occlusion. The `00' and `16' lighting conditions are the most extreme variation from the reference image while the `04' and `12' lighting conditions are much less extreme.  The `Facial Expression' dataset uses an image of the subject with a neutral expression as the reference image then comparison images with the subject either smiling, showing surprise, squinting, showing disgust, or screaming.

\section{Continuous Verification Performance}
\label{appendix:cvperf}

We provide a summary of the FNMR performance of the facial recognition classifiers across sub-sampled datasets we create in order to test the classifier's performance in a continuous verification setting.

\subsection{Average FNMR}
\begin{center}
  \input{tables/continuousVerificationPerformanceSummary}
\end{center}

This table provides a summary of the overall FNMR performance across all of the sub-sampled datasets we created. We see the `Session', `Lighting', and `Facial Expression' datasets exhibiting significantly reduced average FNMR values compared to the `Facial Rotation' dataset.  This is likely due to the fact that the average number of facial features being occluded by any of these variations is significantly less than when the subject rotates their head away from being centered with the camera.  When the facial recognition classifier is unable to find these features they inflate the FNMR since the classifier is calculating the distance between features it locates on the face then comparing the distance to a set threshold in order to determine if a face is verified.

\subsection{Facial Rotation}
\begin{center}
  \input{tables/fnmrRotationPerformance}
\end{center}

This table provides the complete FNMR performance for each facial rotation we tested across all of the different classifiers.  We see the FNMR increase as the facial rotation increases most likely due to more facial features being hidden in the image.  We see FaceNet and OpenFace begin failing to verify the majority of subjects at 60 degrees of facial rotation.  ArcFace begins to fail to verify subjects at 45 degrees of facial rotation while VGG-Face has its performance slowly fall off as the angle increases but it never begins failing to verify a majority of the subjects.  FaceAPI starts failing to verify the majority of subjects at around 90 degrees.  While these face rotations may seem high, it is important to remember even a relatively small FNMR can lead to a significant number of students being flagged for cheating as they take these very high stakes exams.

\subsection{Session}
\begin{center}
  \input{tables/fnmrSessionPerformance}
\end{center}

This table provides the FNMR performance of the classifiers across images taken in different capture sessions.  The session gap refers to the number of sessions between the reference image and the comparison image.  We would expect to see the FNMR increase as the session gap increased but we do not see that pattern with any of the classifiers.  This suggests that the time between captures is less important than other factors such as the subject's facial hair or hairstyle.

\subsection{Lighting}
\begin{center}
  \input{tables/fnmrLightingPerformance}
\end{center}

This table provides the FNMR performance of the different classifiers across the various lighting conditions we test.  We see the FNMR performance degrade when compared against one of the worst lighting conditions, `00'.  However, we do not see a similar degradation when testing against the mirrored condition, `16', suggesting that the performance in bad lighting conditions is very variable across most of the classifiers.  VGG-Face is the only classifier we tested which performed consistently well across all of the lighting conditions.

\end{document}

%% file: tables/lawSchoolAdoption.tex
\begin{table}[]
\centering
\begin{tabular}{@{}lrr@{}}
\toprule
\textbf{Exam Proctoring Suite} & \textbf{Schools} & \textbf{Percentage} \\ \midrule
Examplify                      & 99               & 55\%                \\
Exam4                          & 52               & 29\%                \\
Electronic Blue Book           & 13               & 7.2\%               \\
Canvas                         & 4                & 2.2\%               \\
MyLaw                          & 3                & 1.7\%               \\
ILG Exam360                    & 1                & 0.56\%              \\
Other                          & 5                & 4.5\%               \\ \bottomrule
\end{tabular}
\caption[Proctoring Suite Market Share in Top 180 U.S.Law Schools]{Examplify leads in market share followed by Exam4 and Electronic Blue Book.  Two schools did not respond to inquiries and one closed down.}
\label{tab:lawSchoolAdoption}
\end{table}

%% file: figures/barExamAdoption.tex
\begin{figure}[tp]
    \includegraphics[width=.48\textwidth]{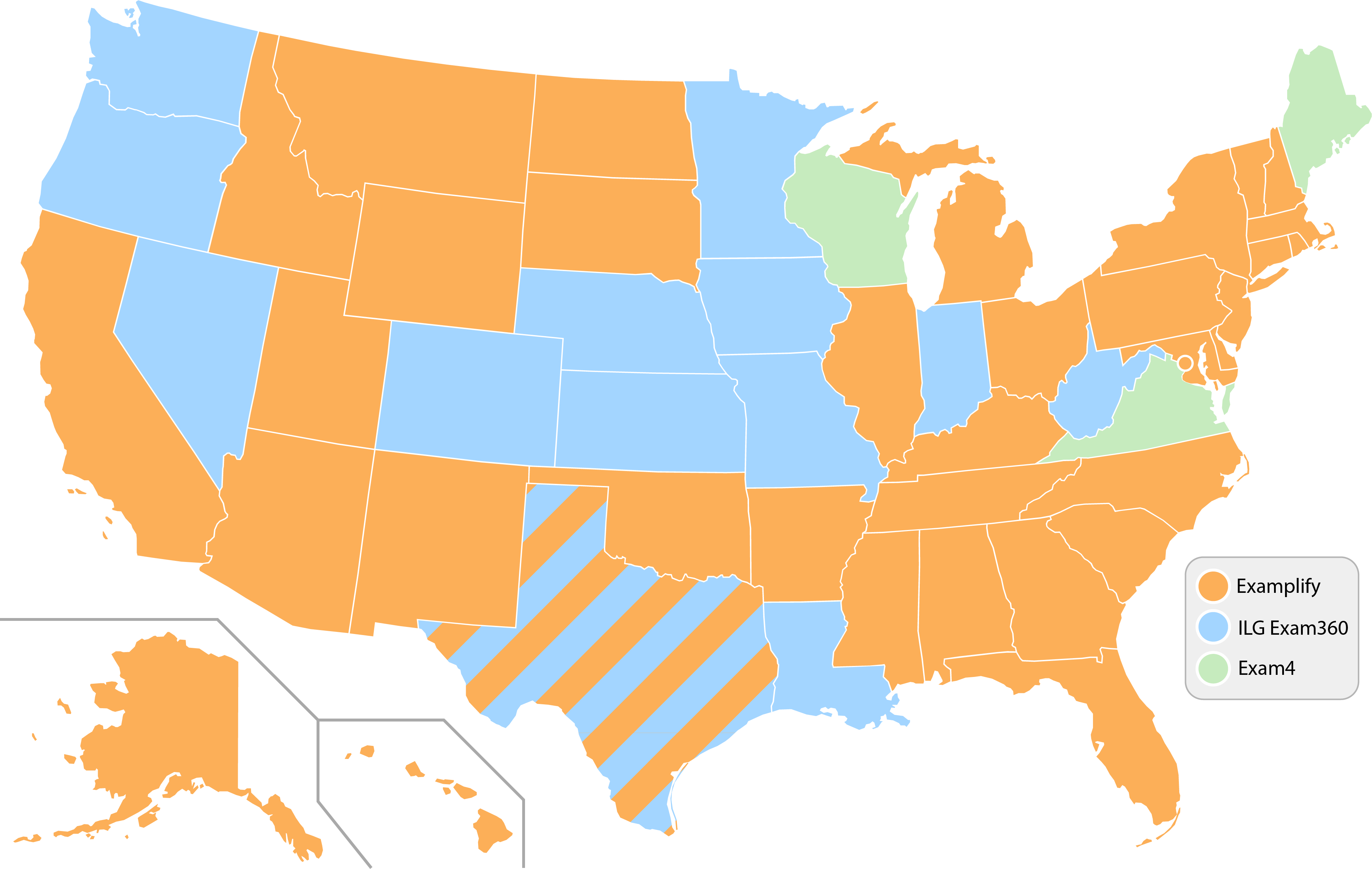}
    \caption[Proctoring Suite Market Share Across Bar Exams]{The adoption of remote exam proctoring suites by state bar exam associations across the United States.}
    \label{fig:barexamadoption}
\end{figure}

%% file: tables/datasetSummary.tex
\begin{table*}[]
\centering
\tabcolsep=4pt\relax
\begin{tabular}{@{}lcrrcccccc@{}}
\toprule
\multicolumn{1}{c}{} &
   &
  \multicolumn{1}{c}{} &
  \multicolumn{1}{c}{} &
   &
  \multicolumn{5}{c}{\textbf{Race Annotation}} \\ \cmidrule(l){6-10} 
\textbf{Dataset} &
  \textbf{Source} &
  \multicolumn{1}{c}{\textbf{\# of subjects}} &
  \multicolumn{1}{c}{\textbf{\# of faces}} &
  \textbf{In-the-wild} &
  White &
  Asian &
  Black &
  Hispanic &
  Indian \\ \midrule
MORPH-II &
  Government Data &
  14K &
  55K &
   &
  \checkmark &
  \checkmark &
  \checkmark &
  \checkmark &
   \\
Multi-PIE &
  Study Capture &
  750K &
  337 &
   &
  \checkmark &
  \checkmark &
  \checkmark &
  \multicolumn{1}{l}{} &
  \multicolumn{1}{l}{} \\
LFWA+ &
  LFW &
  6K &
  13K &
  \checkmark &
  \checkmark &
  \checkmark &
  \checkmark &
   &
  \checkmark \\
VGGFace2 &
  Image Search &
  9K &
  3M &
  \checkmark &
  \checkmark\footnotemark[2] &
  \checkmark\footnotemark[2] &
  \checkmark\footnotemark[2] &
   &
  \checkmark\footnotemark[2] \\ \bottomrule
\end{tabular}
\caption{Summary of datasets used for the facial recognition accuracy section.}
\label{tab:datasets}

\end{table*}

%% file: tables/fnmrFmrSummary.tex
\begin{table*}[]
\centering
\begin{tabular}{@{}llrrlrrlrrlrr@{}}
\toprule
 &
   &
  \multicolumn{2}{c}{\textbf{MORPH-II}} &
   &
  \multicolumn{2}{c}{\textbf{Multi-PIE}} &
   &
  \multicolumn{2}{c}{\textbf{LFWA+}} &
   &
  \multicolumn{2}{c}{\textbf{VGGFace2}} \\ \cmidrule(lr){3-4} \cmidrule(lr){6-7} \cmidrule(lr){9-10} \cmidrule(l){12-13} 
Classifier &
   &
  \multicolumn{1}{c}{FNMR} &
  \multicolumn{1}{c}{FMR} &
   &
  \multicolumn{1}{c}{FNMR} &
  \multicolumn{1}{c}{FMR} &
   &
  \multicolumn{1}{c}{FNMR} &
  \multicolumn{1}{c}{FMR} &
   &
  \multicolumn{1}{c}{FNMR} &
  \multicolumn{1}{c}{FMR} \\ \cmidrule(r){1-1} \cmidrule(lr){3-4} \cmidrule(lr){6-7} \cmidrule(lr){9-10} \cmidrule(l){12-13} 
\textbf{Facenet}  &  & 7.5\%  & 14\%  &  & 9.2\% & 50\% &  & 14\%  & 2.9\%  &  & 17\%  & 2.0\%  \\
\textbf{VGG-Face} &  & 4.1\%  & 78\%  &  & 1.9\% & 83\% &  & 2.8\% & 68\%   &  & 3.4\% & 42\%   \\
\textbf{OpenFace} &  & 5.7\%  & 72\%  &  & 8.6\% & 77\% &  & 12\%  & 62\%   &  & 11\%  & 63\%   \\
\textbf{ArcFace}  &  & 12\%   & 2.2\% &  & 11\%  & 48\% &  & 24\%  & 0.54\% &  & 23\%  & 1.2\%  \\
\textbf{FaceAPI}  &  & 0.66\% & 21\%  &  & 9.2\% & 19\% &  & 22\%  & 0.69\% &  & 14\%  & 0.42\% \\ \bottomrule
\end{tabular}
\caption{FNMR and FMR averages across various datasets using both state of the art algorithms and Face-API.js}
\label{tab:fnmrfmrtable}
\end{table*}

%% file: tables/fnmrFacialRotation.tex
\begin{figure}
\centering
\begin{tikzpicture}

\definecolor{color0}{rgb}{0.12156862745098,0.466666666666667,0.705882352941177}
\definecolor{color1}{rgb}{1,0.498039215686275,0.0549019607843137}
\definecolor{color2}{rgb}{0.172549019607843,0.627450980392157,0.172549019607843}
\definecolor{color3}{rgb}{0.83921568627451,0.152941176470588,0.156862745098039}
\definecolor{color4}{rgb}{0.580392156862745,0.403921568627451,0.741176470588235}

\begin{axis}[
legend cell align={left},
legend style={
  fill opacity=0.8,
  draw opacity=1,
  text opacity=1,
  at={(0.5,0.91)},
  anchor=north,
  draw=white!80!black
},
tick align=outside,
tick pos=left,
x grid style={white!69.0196078431373!black},
xlabel={Facial Tilt (\textdegree{})},
xtick={-90,-75,-60,-45,-30,-15,0,15,30,45,60,75,90},
xmajorgrids,
xmin=-99, xmax=99,
xtick style={color=black},
y grid style={white!69.0196078431373!black},
ylabel={False Non-Match Rate (\%)},
ymajorgrids,
ymin=0, ymax=100,
tick label style={font=\scriptsize},
label style={font=\footnotesize},
legend style={font=\footnotesize},
ytick style={color=black},
ylabel shift = -7 pt
]
\addplot [semithick, color0]
table {%
-90 72.633
-75 72.387
-60 69.189
-45 44.598
-30 21.384
-15 8.974
0 0
15 7.184
30 19.101
45 46.881
60 69.045
75 73.087
90 73.023
};
\addlegendentry{Facenet}
\addplot [semithick, color1]
table {%
-90 34.726
-75 49.356
-60 44.002
-45 17.319
-30 6.205
-15 4.495
0 0
15 3.166
30 4.964
45 22.212
60 40.986
75 36.651
90 44.749
};
\addlegendentry{VGG-Face}
\addplot [semithick, color2]
table {%
-90 71.758
-75 50.366
-60 57.391
-45 36.046
-30 10.557
-15 5.02
0 0
15 4.24
30 9.889
45 42.466
60 49.547
75 63.325
90 72.633
};
\addlegendentry{OpenFace}
\addplot [semithick, color3]
table {%
-90 90.724
-75 89.976
-60 85.394
-45 52.705
-30 19.125
-15 9.857
0 0
15 7.462
30 19.204
45 58.075
60 85.147
75 89.865
90 90.692
};
\addlegendentry{ArcFace}
\addplot [semithick, color4]
table {%
-90 79.785
-75 31.766
-60 7.303
-45 1.615
-30 2.761
-15 0.843
0 0
15 1.249
30 5.8
45 2.8
60 9.475
75 42.259
90 79.268
};
\addlegendentry{FaceAPI}
\end{axis}

\end{tikzpicture}
\caption{False Non Match Rate (FNMR) of subjects at various facial rotations ranging from -90\textdegree{} to 90\textdegree{} from the camera across various classifiers..}
\label{fig:facialrotation}
\end{figure}
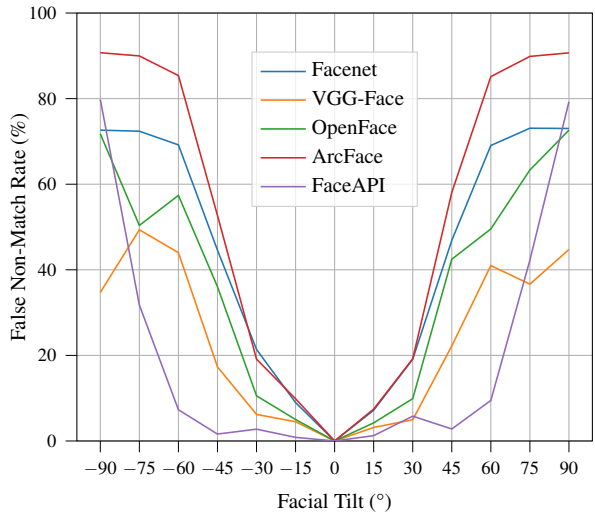

%% file: tables/fnmrRace.tex
\begin{table*}[]
\tabcolsep=4.5pt\relax
\centering
\begin{tabular}{@{}llrrrrlrrrrlrrrr@{}}
\toprule
                  &  & \multicolumn{4}{c}{\textbf{MORPH-II}} &  & \multicolumn{4}{c}{\textbf{LFWA+}} &  & \multicolumn{4}{c}{\textbf{VGGFace2}} \\ \cmidrule(lr){3-6} \cmidrule(lr){8-11} \cmidrule(l){13-16} 
Classifier &
   &
  \multicolumn{1}{c}{\rotatebox[origin=c]{90}{Black}} &
  \multicolumn{1}{c}{\rotatebox[origin=c]{90}{White}} &
  \multicolumn{1}{c}{\rotatebox[origin=c]{90}{Asian}} &
  \multicolumn{1}{c}{\rotatebox[origin=c]{90}{Hispanic}} &
   &
  \multicolumn{1}{c}{\rotatebox[origin=c]{90}{Black}} &
  \multicolumn{1}{c}{\rotatebox[origin=c]{90}{White}} &
  \multicolumn{1}{c}{\rotatebox[origin=c]{90}{Asian}} &
  \multicolumn{1}{c}{\rotatebox[origin=c]{90}{Indian}} &
   &
  \multicolumn{1}{c}{\rotatebox[origin=c]{90}{Black}} &
  \multicolumn{1}{c}{\rotatebox[origin=c]{90}{White}} &
  \multicolumn{1}{c}{\rotatebox[origin=c]{90}{Asian}} &
  \multicolumn{1}{c}{\rotatebox[origin=c]{90}{Indian}} \\ \cmidrule(r){1-1} \cmidrule(lr){3-6} \cmidrule(lr){8-11} \cmidrule(l){13-16} 
\textbf{Facenet}  &  & 6.8\%    & 9.7\%   & 15\%    & 10\%   &  & 15\%   & 15\%   & 9.5\%  & 14\%    &  & 26\%    & 17\%    & 17\%    & 15\%    \\
\textbf{VGG-Face} &  & 3.6\%    & 5.2\%   & 11\%    & 7.4\%  &  & 1.4\%  & 2.9\%  & 2.7\%  & 0.58\%  &  & 1.9\%   & 3.6\%   & 3.1\%   & 3.4\%   \\
\textbf{OpenFace} &  & 6.0\%    & 4.6\%   & 9.5\%   & 3.8\%  &  & 13\%   & 12\%   & 10\%   & 18\%    &  & 18\%    & 10\%    & 14\%    & 7.1\%   \\
\textbf{ArcFace}  &  & 11\%     & 13\%    & 23\%    & 11\%   &  & 30\%   & 24\%   & 15\%   & 27\%    &  & 34\%    & 21\%    & 27\%    & 19\%    \\
\textbf{FaceAPI}  &  & 0.57\%   & 1.1\%   & 0.0\%   & 0.1\%  &  & 23\%   & 22\%   & 18\%   & 21\%    &  & 20\%    & 13\%    & 14\%    & 11\%    \\ \bottomrule
\end{tabular}
\caption{FNMR averages across various datasets using both state of the art algorithms and Face-API.js}
\label{tab:racefnmr}
\end{table*}

%% file: tables/securityFeatureComparison.tex
{
\tabcolsep=4.5pt\relax
\centering
\begin{tabular}{@{}llcccc@{}}
\toprule
Security Feature                     &  & \textbf{Examplify} & \textbf{Exam4}  & \textbf{EBB} & \textbf{ILG Exam360} \\ \cmidrule(r){1-1} \cmidrule(l){3-6} 
\textbf{Encryption at Rest}       &  & AES-256    & AES-256    & 3DES    & AES-256    \\
\textbf{Encryption in Transit}    &  & HTTPS      & HTTP       & HTTP    & HTTPS      \\
\textbf{Virtual Machine Protection}  &  & Block List         & Block List      & Block List   & Block List           \\
\textbf{Virtual Device Detection} &  & Block List & N/I        & N/I     & Block List \\
\textbf{Clipboard Management}     &  & Integrated & Cleared    & Cleared & Integrated \\
\textbf{Screenshot Capture}       &  & N/I        & N/I        & N/I     & App Window \\
\textbf{Process Restrictions}     &  & Allow List & Block List & N/I     & Null DNS   \\
\textbf{Network Access Restrictions} &  & Route Table        & Adapter Disable & N/I          & Null DNS             \\ \bottomrule
\end{tabular}
}

%% file: tables/privacyFeatureComparison.tex
{
\tabcolsep=4.5pt\relax
\centering
\begin{tabular}{@{}llcccc@{}}
\toprule
Privacy Related Feature &  & \textbf{Examplify} & \textbf{Exam4} & \textbf{EBB} & \textbf{ILG Exam360} \\ \cmidrule(r){1-1} \cmidrule(l){3-6} 
\textbf{Initial Identity Verification} &  & Automated            & N/I         & N/I         & Human       \\
\textbf{Continuous Identity Check}     &  & Automated            & N/I         & N/I         & Human       \\
\textbf{System Service}                &  & Always Running       & App Running & App Running & App Running \\
\textbf{Device Identifiers}            &  & App List/OS/Hardware & N/I         & N/I         & OS/Hardware \\ \bottomrule
\end{tabular}
}

%% file: tables/virtualMachineDetection.tex
{
\centering
\begin{tabular}{@{}llcccc@{}}
\toprule
Virtual Machine             &  & \textbf{Examplify} & \textbf{Exam4} & \textbf{EBB} & \textbf{ILG Exam360} \\ \cmidrule(r){1-1} \cmidrule(l){3-6} 
\textbf{VirtualBox}         &  & \checkmark              & \checkmark          & \checkmark        & \checkmark                \\
\textbf{VMWare Workstation} &  & \checkmark              & \checkmark          & \checkmark        & \checkmark                \\
\textbf{VMWare Fusion}      &  & \checkmark              &                &              &                      \\
\textbf{Parallels}          &  & \checkmark              & \checkmark          &              & \checkmark                \\
\textbf{Hyper-V}            &  & \checkmark              &                &              &                      \\
\textbf{QEMU}               &  & \checkmark              &                &              &                      \\ \bottomrule
\end{tabular}
}

%% file: tables/virtualDeviceDetection.tex
{
\centering
\begin{tabular}{@{}llcccc@{}}
        \toprule
        Virtual Webcam/Microphone &  & \textbf{Examplify} & \textbf{Exam4} & \textbf{EBB} & \textbf{ILG Exam360} \\ \cmidrule(r){1-1} \cmidrule(l){3-6} 
        \textbf{ManyCam}          &  & \checkmark & N/I & N/I & \checkmark \\
        \textbf{YouCam}           &  & \checkmark & N/I & N/I & \checkmark \\
        \textbf{MyCam}            &  & \checkmark & N/I & N/I & \checkmark \\
        \textbf{Logitech Capture} &  &       & N/I & N/I &       \\
        \textbf{OBS Studio}       &  &       & N/I & N/I & \checkmark \\ \bottomrule
\end{tabular}
}

%% file: tables/initialVerificationDatasetSummary.tex

\begin{tabular}{@{}llcclcc@{}}
\toprule
                   &  & \multicolumn{2}{c}{\textbf{FNMR}}  &  & \multicolumn{2}{c}{\textbf{FMR}}   \\ \cmidrule(lr){3-4} \cmidrule(l){6-7} 
Classifier         &  & \# of subjects & \# of comparisons &  & \# of subjects & \# of comparisons \\ \cmidrule(r){1-1} \cmidrule(lr){3-4} \cmidrule(l){6-7} 
\textbf{MORPH-II}  &  & 13K            & 42K               &  & 14K            & 681K              \\
\textbf{Multi-PIE} &  & 264            & 81K               &  & 264            & 172K              \\
\textbf{LFWA+}     &  & 1.7K           & 7.4K              &  & 5.7K           & 287K              \\
\textbf{VGGFace2}  &  & 9.1K           & 457K              &  & 9.1K           & 457K              \\ \bottomrule
\end{tabular}

%% file: tables/continuousVerificationDatasetSummary.tex
{
\tabcolsep=2pt\relax
\centering
\begin{tabular}{@{}llcc@{}}
\toprule
Multi-PIE Dataset          &  & \# of subjects & \# of comparisons \\ \cmidrule(r){1-1} \cmidrule(l){3-4} 
\textbf{Session}           &  & 264            & 81K               \\
\textbf{Facial Rotation}       &  & 337            & 151K              \\
\textbf{Lighting}          &  & 337            & 131K              \\
\textbf{Facial Expression} &  & 337            & 104K              \\ \bottomrule
\end{tabular}
}

%% file: tables/continuousVerificationPerformanceSummary.tex
{
\centering
\begin{tabular}{@{}llrrrr@{}}
\toprule
                  &  & \multicolumn{4}{c}{\textbf{Multi-PIE Dataset}} \\ \cmidrule(l){3-6} 
Classifier &  & \multicolumn{1}{c}{Session} & \multicolumn{1}{c}{Facial Rotation} & \multicolumn{1}{c}{Lighting} & \multicolumn{1}{c}{Facial Expression} \\ \cmidrule(r){1-1} \cmidrule(l){3-6} 
\textbf{Facenet}  &  & 9.2\%      & 48\%      & 20\%      & 9.4\%     \\
\textbf{VGG-Face} &  & 1.9\%      & 26\%      & 2.3\%     & 2.1\%     \\
\textbf{OpenFace} &  & 8.6\%      & 39\%      & 11\%      & 7.7\%     \\
\textbf{ArcFace}  &  & 11\%       & 58\%      & 14\%      & 11\%      \\
\textbf{FaceAPI}  &  & 9.2\%      & 22\%      & 8.7\%     & 9.6\%     \\ \bottomrule
\end{tabular}
}

%% file: tables/fnmrRotationPerformance.tex
{
\centering
\begin{tabular}{@{}llrrrrrrrrrrrr@{}}
\toprule
                  &  & \multicolumn{12}{c}{\textbf{Rotation}}                                                        \\ \cmidrule(l){3-14} 
Classifier &
   &
  \multicolumn{1}{c}{-90} &
  \multicolumn{1}{c}{-75} &
  \multicolumn{1}{c}{-60} &
  \multicolumn{1}{c}{-45} &
  \multicolumn{1}{c}{-30} &
  \multicolumn{1}{c}{-15} &
  \multicolumn{1}{c}{15} &
  \multicolumn{1}{c}{30} &
  \multicolumn{1}{c}{45} &
  \multicolumn{1}{c}{60} &
  \multicolumn{1}{c}{75} &
  \multicolumn{1}{c}{90} \\ \cmidrule(r){1-1} \cmidrule(l){3-14} 
\textbf{Facenet}  &  & 73\% & 72\% & 69\%  & 45\%  & 21\%  & 9.0\%  & 7.2\% & 19\%  & 47\%  & 69\%  & 73\% & 73\% \\
\textbf{VGG-Face} &  & 35\% & 49\% & 44\%  & 17\%  & 6.2\% & 4.5\%  & 3.2\% & 5.0\% & 22\%  & 41\%  & 37\% & 45\% \\
\textbf{OpenFace} &  & 72\% & 50\% & 57\%  & 36\%  & 11\%  & 5.0\%  & 4.2\% & 9.9\% & 42\%  & 50\%  & 63\% & 73\% \\
\textbf{ArcFace}  &  & 91\% & 90\% & 85\%  & 53\%  & 19\%  & 9.9\%  & 7.4\% & 19\%  & 58\%  & 85\%  & 90\% & 91\% \\
\textbf{FaceAPI}  &  & 80\% & 32\% & 7.3\% & 1.6\% & 2.8\% & 0.84\% & 1.2\% & 5.8\% & 2.8\% & 9.5\% & 42\% & 79\% \\ \bottomrule
\end{tabular}
}

%% file: tables/fnmrSessionPerformance.tex
{
\centering
\begin{tabular}{@{}llrrr@{}}
\toprule
                  &  & \multicolumn{3}{c}{\textbf{Session Gap}}                              \\ \cmidrule(l){3-5} 
Classifier        &  & \multicolumn{1}{c}{1} & \multicolumn{1}{c}{2} & \multicolumn{1}{c}{3} \\ \cmidrule(r){1-1} \cmidrule(l){3-5} 
\textbf{Facenet}  &  & 10\%                  & 9.2\%                 & 8.0\%                 \\
\textbf{VGG-Face} &  & 2.1\%                 & 1.8\%                 & 1.8\%                 \\
\textbf{OpenFace} &  & 9.3\%                 & 8.4\%                 & 7.7\%                 \\
\textbf{ArcFace}  &  & 12\%                  & 10\%                  & 9.2\%                 \\
\textbf{FaceAPI}  &  & 10\%                  & 8.9\%                 & 8.3\%                 \\ \bottomrule
\end{tabular}
}

%% file: tables/fnmrLightingPerformance.tex
{
\centering
\begin{tabular}{@{}llrrrr@{}}
\toprule
                  &  & \multicolumn{4}{c}{\textbf{Lighting}} \\ \cmidrule(l){3-6} 
Classifier &  & \multicolumn{1}{c}{00} & \multicolumn{1}{c}{04} & \multicolumn{1}{c}{12} & \multicolumn{1}{c}{16} \\ \cmidrule(r){1-1} \cmidrule(l){3-6} 
\textbf{Facenet}  &  & 34\%    & 15\%    & 17\%    & 13\%    \\
\textbf{VGG-Face} &  & 1.1\%   & 2.7\%   & 2.4\%   & 3.0\%   \\
\textbf{OpenFace} &  & 13\%    & 12\%    & 12\%    & 8.4\%   \\
\textbf{ArcFace}  &  & 15\%    & 15\%    & 15\%    & 13\%    \\
\textbf{FaceAPI}  &  & 17\%    & 7.2\%   & 8.2\%   & 2.9\%   \\ \bottomrule
\end{tabular}
}

%% file: paper.bib
@misc{parkhi2015deep,
  title={Deep face recognition. University of Oxford},
  author={Parkhi, OM and Vedaldi, A and Zisserman, A},
  year={2015},
  publisher={Oxford, United Kingdom}
}

@article{Greco_MVA2020,
        author="Greco, Antonio and Percannella, Gennaro and Vento, Mario and Vigilante, Vincenzo",
        title="Benchmarking deep network architectures for ethnicity recognition using a new large face dataset",
        journal="Machine Vision and Applications",
        year="2020",
        publisher="Springer"
}

@misc{cao2018vggface2,
      title={VGGFace2: A dataset for recognising faces across pose and age}, 
      author={Qiong Cao and Li Shen and Weidi Xie and Omkar M. Parkhi and Andrew Zisserman},
      year={2018},
      eprint={1710.08092},
      archivePrefix={arXiv},
      primaryClass={cs.CV}
}

@inproceedings{liu2015faceattributes,
 author = {Liu, Ziwei and Luo, Ping and Wang, Xiaogang and Tang, Xiaoou},
 title = {Deep Learning Face Attributes in the Wild},
 booktitle = {Proceedings of International Conference on Computer Vision (ICCV)},
 year = {2015} 
}

@phdthesis{cheng2016binary,
  title={Binary Analysis and Symbolic Execution with angr},
  author={Cheng, Eric},
  year={2016},
  school={PhD thesis, The MITRE Corporation}
}

@inproceedings{serengil2020lightface,
  title={LightFace: A Hybrid Deep Face Recognition Framework},
  author={Serengil, Sefik Ilkin and Ozpinar, Alper},
  booktitle={2020 Innovations in Intelligent Systems and Applications Conference (ASYU)},
  pages={23-27},
  year={2020},
  doi={10.1109/ASYU50717.2020.9259802},
  organization={IEEE}
}

@inproceedings{deng2019retinaface,
title={RetinaFace: Single-stage Dense Face Localisation in the Wild},
author={Deng, Jiankang and Guo, Jia and Yuxiang, Zhou and Jinke Yu and Irene Kotsia and Zafeiriou, Stefanos},
booktitle={arxiv},
year={2019}
}

@techreport{amos2016openface,
  title={OpenFace: A general-purpose face recognition library with mobile applications},
  author={Amos, Brandon and Bartosz, Ludwiczuk and Satyanarayanan, Mahadev},
  year={2016},
  institution={CMU-CS-16-118, CMU School of Computer Science},
}

@article{barrett,
author = {Barrett, Lindsey},
year = {2021},
month = {01},
pages = {},
title = {Rejecting Test Surveillance in Higher Education},
journal = {SSRN Electronic Journal},
doi = {10.2139/ssrn.3871423}
}

@misc{inteldatasheet, title={12th Generation Intel® Core™ Processors Datasheet}, url={https://cdrdv2.intel.com/v1/dl/getcontent/655258}, author={Corporation, Intel}}

@inproceedings{guo2018stacked,
  title={Stacked Dense U-Nets with Dual Transformers for Robust Face Alignment},
  author={Guo, Jia and Deng, Jiankang and Xue, Niannan and Zafeiriou, Stefanos},
  booktitle={BMVC},
  year={2018}
}

@article{deng2018menpo,
  title={The Menpo benchmark for multi-pose 2D and 3D facial landmark localisation and tracking},
  author={Deng, Jiankang and Roussos, Anastasios and Chrysos, Grigorios and Ververas, Evangelos and Kotsia, Irene and Shen, Jie and Zafeiriou, Stefanos},
  journal={IJCV},
  year={2018}
}

@inproceedings{deng2018arcface,
title={ArcFace: Additive Angular Margin Loss for Deep Face Recognition},
author={Deng, Jiankang and Guo, Jia and Niannan, Xue and Zafeiriou, Stefanos},
booktitle={CVPR},
year={2019}
}

@inproceedings{schroff2015facenet,
  title={Facenet: A unified embedding for face recognition and clustering},
  author={Schroff, Florian and Kalenichenko, Dmitry and Philbin, James},
  booktitle={Proceedings of the IEEE conference on computer vision and pattern recognition},
  pages={815--823},
  year={2015}
}

@inproceedings{morph,
author = {Ricanek, Karl and Tesafaye, T.},
year = {2006},
month = {05},
pages = {341 - 345},
title = {MORPH: A longitudinal image database of normal adult age-progression},
volume = {2006},
isbn = {0-7695-2503-2},
journal = {FGR 2006: Proceedings of the 7th International Conference on Automatic Face and Gesture Recognition},
doi = {10.1109/FGR.2006.78}
}

@article{gross2010multi,
  title={Multi-pie},
  author={Gross, Ralph and Matthews, Iain and Cohn, Jeffrey and Kanade, Takeo and Baker, Simon},
  journal={Image and vision computing},
  volume={28},
  number={5},
  pages={807--813},
  year={2010},
  publisher={Elsevier}
}

@article{tefft2014driver,
  title={Driver license renewal policies and fatal crash involvement rates of older drivers, United States, 1986--2011},
  author={Tefft, Brian C},
  journal={Injury epidemiology},
  volume={1},
  number={1},
  pages={1--11},
  year={2014},
  publisher={BioMed Central}
}

@article{slusky2020cybersecurity,
  title={Cybersecurity of Online Proctoring Systems},
  author={Slusky, Ludwig},
  journal={Journal of International Technology and Information Management},
  volume={29},
  number={1},
  pages={56--83},
  year={2020}
}

@inproceedings{wu2020gender,
  title={Gender classification and bias mitigation in facial images},
  author={Wu, Wenying and Protopapas, Pavlos and Yang, Zheng and Michalatos, Panagiotis},
  booktitle={12th ACM Conference on Web Science},
  pages={106--114},
  year={2020}
}

@article{karim2014cheating,
  title={Cheating, reactions, and performance in remotely proctored testing: An exploratory experimental study},
  author={Karim, Michael N and Kaminsky, Samuel E and Behrend, Tara S},
  journal={Journal of Business and Psychology},
  volume={29},
  number={4},
  pages={555--572},
  year={2014},
  publisher={Springer}
}

@article{teclehaimanot2018ensuring,
  title={Ensuring academic integrity in online courses: A case analysis in three testing environments},
  author={Teclehaimanot, Berhane and You, Jiyu and Franz, Diana R and Xiao, Mingli and Hochberg, Sue Ann},
  journal={The Quarterly Review of Distance Education},
  volume={12},
  number={1},
  pages={47--52},
  year={2018}
}

@inproceedings{turani2020students,
  title={Students Online Exam Proctoring: A Case Study Using 360 Degree Security Cameras},
  author={Turani, Aiman A and Alkhateeb, Jawad H and Alsewari, AbdulRahman A},
  booktitle={2020 Emerging Technology in Computing, Communication and Electronics (ETCCE)},
  pages={1--5},
  year={2020},
  organization={IEEE}
}

@article{langenfeld2020internet,
  title={Internet-Based Proctored Assessment: Security and Fairness Issues},
  author={Langenfeld, Thomas},
  journal={Educational Measurement: Issues and Practice},
  volume={39},
  number={3},
  pages={24--27},
  year={2020},
  publisher={Wiley Online Library}
}

@article{leslie2020understanding,
  title={Understanding bias in facial recognition technologies},
  author={Leslie, David},
  journal={arXiv preprint arXiv:2010.07023},
  year={2020}
}

@book{grother2014face,
  title={Face recognition vendor test (frvt)},
  author={Grother, Patrick J and Grother, Patrick J and Ngan, Mei},
  year={2014},
  publisher={US Department of Commerce, National Institute of Standards and Technology}
}

@article{nagpal2019deep,
  title={Deep learning for face recognition: Pride or prejudiced?},
  author={Nagpal, Shruti and Singh, Maneet and Singh, Richa and Vatsa, Mayank},
  journal={arXiv preprint arXiv:1904.01219},
  year={2019}
}

@inproceedings{singh2020robustness,
  title={On the robustness of face recognition algorithms against attacks and bias},
  author={Singh, Richa and Agarwal, Akshay and Singh, Maneet and Nagpal, Shruti and Vatsa, Mayank},
  booktitle={Proceedings of the AAAI Conference on Artificial Intelligence},
  volume={34},
  number={09},
  pages={13583--13589},
  year={2020}
}

@misc{ars-antivirus,
 title={It might be time to stop using antivirus},
 note={\url{https://arstechnica.com/information-technology/2017/01/antivirus-is-bad/}},
 author={Anthony, Sebastian},
 howpublished={Ars Technica},
 year={2017}
}

@article{raman2021adoption,
  title={Adoption of online proctored examinations by university students during COVID-19: Innovation diffusion study},
  author={Raman, Raghu and Sairam, B and Veena, G and Vachharajani, Hardik and Nedungadi, Prema},
  journal={Education and Information Technologies},
  pages={1--20},
  year={2021},
  publisher={Springer}
}

@article{feathers, 
title={Proctorio Is Using Racist Algorithms to Detect Faces}, 
note={\url{https://www.vice.com/en/article/g5gxg3/proctorio-is-using-racist-algorithms-to-detect-faces}}, journal={VICE}, author={Feathers, Todd},
year={2021}

}

@article{studenttweet,
title={Hey @proctorio @artfulhacker How do you explain this?},
year={2020},
note={\url{http://web.archive.org/web/20210715164139/https://twitter.com/ejohnson99/status/1303121786637373443}},
author={Johnson, Erik},
journal={Twitter}
}

@article{techrev,
author={Swauger, Shea},
note={\url{https://www.technologyreview.com/2020/08/07/1006132/software-algorithms-proctoring-online-tests-ai-ethics/}},
journal={MIT Technology Review},
year={2020}
}

@article{balash-proctoring,
    title={{Examining the Examiners: Students’ Privacy and Security Perceptions of Online Proctoring Services}},
    author={Balash, David G. and Kim, Dongkun and Shaibekova, Darika and Fainchtein, Rahel A. and Sherr, Micah and Aviv, Adam J.},
    year={2021},
    journal={USENIX Symposium on Usable Privacy and Security}
}

@article{cohney2020virtual,
  title={Virtual Classrooms and Real Harms},
  author={Cohney, Shaanan and Teixeira, Ross and Kohlbrenner, Anne and Narayanan, Arvind and Kshirsagar, Mihir and Shvartzshnaider, Yan and Sanfilippo, Madelyn},
  journal={USENIX Symposium on Usable Privacy and Security},
  year={2021}
}

@misc{bloombergbias,
note={\url{https://news.bloomberglaw.com/health-law-and-business/online-bar-exams-come-with-face-scans-discrimination-concerns}},
title={{Online Bar Exams Come With Face Scans, Bias Concerns}},
author={Reed, Allie},
howpublished={Bloomberg Law},
year={2020}
}

@misc{nytimesdartmouth,
title={{Online Cheating Charges Upend Dartmouth Medical School}},
author={Natasha Singer},
howpublished={NYTimes},
year={2021}
}

@article{appelbaum1987informed,
  title={Informed consent: Legal theory and clinical practice},
  author={Appelbaum, Paul S and Lidz, Charles W and Meisel, Alan},
  year={1987}
}

@inproceedings{kemelmacher2016megaface,
  title={The megaface benchmark: 1 million faces for recognition at scale},
  author={Kemelmacher-Shlizerman, Ira and Seitz, Steven M and Miller, Daniel and Brossard, Evan},
  booktitle={Proceedings of the IEEE conference on computer vision and pattern recognition},
  pages={4873--4882},
  year={2016}
}

@article{prelaw2019,
    title={Be prepared: Law school doesn’t even resemble your college experience}}
